\documentclass[twocolumn,showpacs,preprintnumbers,amsmath,amssymb,prl,superscriptaddress]{revtex4}
\usepackage{mathrsfs}
\usepackage{graphicx}
\usepackage{dcolumn}
\usepackage{bm}
\usepackage{amsmath}
\usepackage{amsfonts}

\begin{document}

\title{Power-Law Decay of Standing Waves on the Surface of Topological Insulators}
\author{Jing Wang}
\affiliation{ State Key Laboratory of Low-Dimensional Quantum Physics, and Department of Physics, Tsinghua University, Beijing 100084,
China}
\author{Wei Li}
\affiliation{ State Key Laboratory of Low-Dimensional Quantum Physics, and Department of Physics, Tsinghua University, Beijing 100084,
China}
\author{Peng Cheng}
\affiliation{ State Key Laboratory of Low-Dimensional Quantum Physics, and Department of Physics, Tsinghua University, Beijing 100084,
China}
\author{Canli Song}
\affiliation{ State Key Laboratory of Low-Dimensional Quantum Physics, and Department of Physics, Tsinghua University, Beijing 100084,
China}
\author{Tong Zhang}
\affiliation{ State Key Laboratory of Low-Dimensional Quantum Physics, and Department of Physics, Tsinghua University, Beijing 100084,
China} \affiliation{Institute of Physics, Chinese Academy of
Sciences, Beijing 100190, China}
\author{Peng Deng}
\affiliation{ State Key Laboratory of Low-Dimensional Quantum Physics, and Department of Physics, Tsinghua University, Beijing 100084,
China}
\author{Xi Chen}
\thanks{xc@mail.tsinghua.edu.cn}
\affiliation{ State Key Laboratory of Low-Dimensional Quantum Physics, and Department of Physics, Tsinghua University, Beijing 100084,
China}
\author{Xucun Ma}
\affiliation{Institute of Physics, Chinese Academy of Sciences, Beijing 100190, China}
\author{Ke He}
\affiliation{Institute of Physics, Chinese Academy of Sciences, Beijing 100190, China}
\author{Jin-Feng Jia}
\affiliation{ State Key Laboratory of Low-Dimensional Quantum Physics, and Department of Physics, Tsinghua University, Beijing 100084,
China}
\author{Qi-Kun Xue}
\affiliation{ State Key Laboratory of Low-Dimensional Quantum Physics, and Department of Physics, Tsinghua University, Beijing 100084,
China} \affiliation{Institute of Physics, Chinese Academy of
Sciences, Beijing 100190, China}
\author{Bang-Fen Zhu}
\thanks{bfz@mail.tsinghua.edu.cn}
\affiliation{ State Key Laboratory of Low-Dimensional Quantum Physics, and Department of Physics, Tsinghua University, Beijing 100084,
China} \affiliation{The Institute of Advanced Study, Tsinghua
University, Beijing 100084, China}

\date{\today}

\begin{abstract}

We propose a general theory on the standing waves (quasiparticle interference pattern)
caused by the scattering of surface states off step edges in topological insulators, in which the extremal points on the constant energy contour of surface band play the dominant role. Experimentally we image the interference patterns on both Bi$_2$Te$_3$ and Bi$_2$Se$_3$ films by measuring the local density of states using a scanning tunneling microscope. The observed decay indices of the standing waves agree excellently with the theoretical prediction: In Bi$_2$Se$_3$, only a single decay index of $-3/2$  exists; while in Bi$_2$Te$_3$ with strongly warped surface band, it varies from $-3/2$  to $-1/2$  and finally to $-1$ as the energy increases. The $-1/2$ decay indicates that the suppression of backscattering due to time-reversal symmetry does not necessarily lead to a spatial decay rate faster than that in the conventional two-dimensional electron system. Our formalism can also explain the characteristic scattering wave vectors of the standing wave caused by non-magnetic impurities on Bi$_2$Te$_3$.

\end{abstract}


\maketitle

The discovery of topological insulators (TIs) has attracted a great
deal of
attention~\cite{qi2010a,moore2010,hasan2010,qi2010b,fu2007,moore2007,hsieh2008,chen2009,zhanghj2009,xia2009,hsieh2009,zhangyi2010}.
The three-dimensional TIs are characterized by the gapped bulk
states and gapless surface states~(SSs), which are protected by
time-reversal symmetry~(TRS) and consist of an odd number of
spin-helical Dirac cones. Exotic effects such as Majorana
fermions~\cite{fu2008,qi2009a} and magnetic
monopole~\cite{qi2009b} are predicted to exist as results of
the topological SSs.

The low-temperature scanning tunneling microscope~(STM) and
spectroscopy~(STS) provide a direct way to study the SSs through
probing the local density of states~(LDOS) oscillations in the
vicinity of impurities or step edges~\cite{crommie1993}. The
quasiparticle interference (QPI) patterns induced by non-magnetic
impurities on the surface of
Bi$_{x}$Sb$_{1-x}$~\cite{roushan2009} and
Bi$_2$Te$_3$~\cite{zhangtong2009}, together with subsequent
theoretical
analysis~\cite{zhou2009,lee2009,wang2010,guo2010,biswas2010,biswas2011},
demonstrated the absence of backscattering for the topological SSs.
Meanwhile, the LDOS oscillations of SSs near step edges on
Bi$_2$Te$_3$ showed a power-law decay with
index $-1$ in a certain energy range~\cite{alpichshev2010},
compared to $-1/2$ for the conventional two-dimensional electron
system~(2DES)~\cite{crommie1993}. The faster decay of QPI once
again indicates the suppression of backscattering in TIs.

Despite the intensive investigation, a complete understanding of QPI
on the surface of TIs remain elusive partially due to the warping
effect of the Dirac cone~\cite{chen2009,fu2009}. The
warping effect of the SSs results from not only the anisotropic
surface band dispersion, but also the coupling between the surface
and bulk bands. In this Letter, we present a general formalism to
account for the complex scattering geometry. We propose the
interference patterns are dominated by the extremal points on the
constant energy contour~(CEC) of 2D electron band. In applying the
theory to Bi$_2$Te$_3$ with strong warping effect, we show that the
decay index varies from $-3/2$ to $-1/2$ and finally to $-1$ as the
energy increases. As for TIs with nearly isotropic Dirac cones, such
as Bi$_2$Se$_3$~\cite{kuroda2010}, the decay index is simply
$-3/2$. Moreover, the theory can be extended to QPI induced by point
defects and readily elucidate the missing of $\vec{q}_3$ and the
deviation of Fermi velocity in Ref.~\cite{zhangtong2009}. To
confirm the predictions, we have performed STM study on both
Bi$_2$Te$_3$ and Bi$_2$Se$_3$ films and found excellent agreement
with theory. In particular, we have been able to obtain the decay
index on Bi$_2$Se$_3$, whose interference pattern is usually too
weak to extract the information.

We start with a general 2D surface band with a single Fermi surface
within the surface Brillouin zone~(SBZ). Because of elastic
scattering, the incoming surface wave with a wave vector
$\mathbf{k}^i$ must be scattered into the outgoing one with
$\mathbf{k}^f$ on the same CEC.  Assuming a step edge along the
$y$-direction, the $k_y$ component of the wave vectors should be
conserved in a scattering process, i.e. $k^i_y=k^f_y\equiv k_y$. The
interference between the incoming and outgoing waves  gives rise to
the standing wave oscillation in the $x$-direction. The total LDOS
is the sum of contributions from all these oscillations from the SSs
on a CEC.  For a given energy $E$, we can integrate over $k_y$ on
the entire CEC and express the LDOS explicitly as
\begin{equation}\label{integral}
\delta\rho(E,x) = \Re\left[\oint_E 2r/(1+\left|r\right|^2)\xi_i^{\dag}\xi_fe^{i(k^f_x-k^i_x)x}dk_y\right],
\end{equation}
where $r$ is the reflection coefficient. Here  the scattering wave function is of the form $\xi e^{ik_xx+ik_yy}$
($\xi$ denotes the spin wave function).

A pair of scattering states $\mathbf{k}^i$ and $\mathbf{k}^f$ lead
to a standing wave with a spatial period of $2\pi/(k^f_x-k^i_x)$.
Since the period is different for different value of $k_y$, only the
pairs  whose periods are \emph{stationary}
with respect to small variation in $k_y$  can make dominant
contribution to the LDOS oscillations. We call such pair of points (with identical $k_y$) on CEC
the \emph{extremal points}~(EPs)~\cite{roth1966}. Other standing
waves interfere with each other and cancel at large $x$. The spatial
dependence of LDOS oscillations in Eq.~(\ref{integral}) can be
evaluated by expanding the relevant quantities around each EP,
namely, let $k_y=k_{y0}+\delta k_y$, then $k_x^f-k_x^i=\Delta k_{x0}
+ \sum_{n}\Delta k_{xn}\delta k_y^{n}$, $r=\sum_{l}\eta_l\delta
k_y^{l}$, and $\xi_i^{\dag}\xi_f=\sum_{m} \chi_m\delta k_y^{m}$.
Here $\Delta k_{x0}$ is the characteristic wave vector depending on
the geometry of CEC. To the leading order of $k_y$, the LDOS varies
at long distance as
\begin{eqnarray}\label{LDOS}
\delta \rho(E,x) &\simeq& \Re\left[\sum\limits_{\mathrm{EPs}}\int_E2r/(1+\left|r\right|^2)\xi_i^{\dag}\xi_fe^{i(k_x^f-k_x^i)x}dk_y\right],
\nonumber
\\
&\sim& \sum\limits_{\mathrm{EPs}}\left|g\eta_{a}\chi_{b}c\right|\cos(\Delta k_{x0}x+\phi_s)x^{-\frac{a+b+1}{c}},
\end{eqnarray}
where $a$=$\min{(l)}$, $b$=$\min{(m)}$, $c$=$\min{(n)}$, and $\phi_s$ is the initial phase of each
 pair of EPs. $g$ is given by
 $\oint_E dk' k'^{(a+b-c+1)/c}e^{i\Delta k_{xc}k'}$.
 The decay behavior of LDOS in Eq.~(\ref{LDOS}) is valid as long as $x\gg\Delta k_{x0}^{-1}$.
 The decay index associated with a pair of EPs is given by $(a+b+1)/c$, which is solely determined by the properties of the scattering wave function around the EPs.

\begin{figure}[t]
\begin{center}
\includegraphics[width=3.3in]{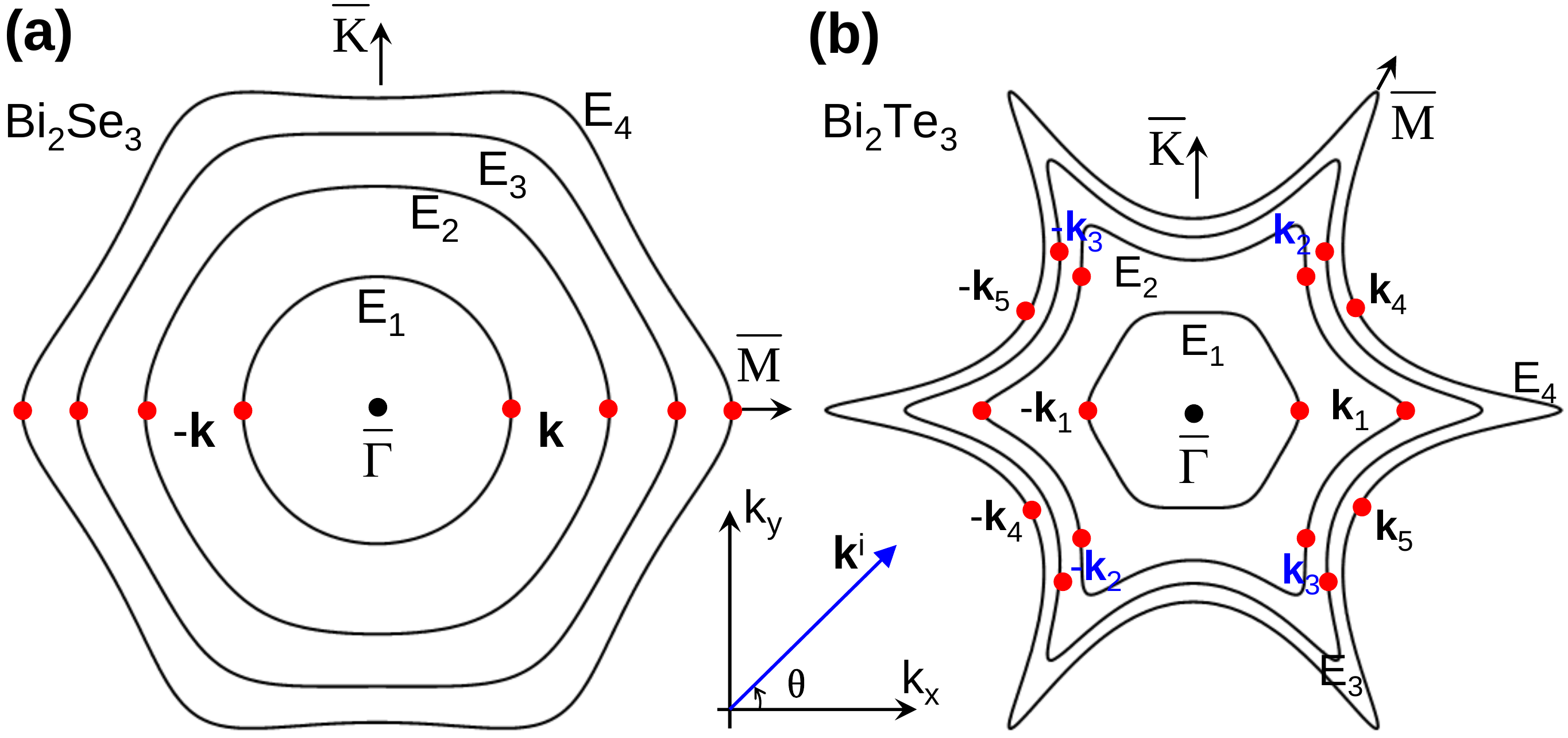}
\end{center}
\caption{(color online) Illustration of scattering off the step edge along $y$-direction. The LDOS oscillation is dominated by scattering between the EP pairs on CEC (red dots). (a) The shape of CEC evolves from circle, hexagon to concave hexagon as energy increases in Bi$_2$Se$_3$. It has a single pair of EPs at ($\mathbf{k}$, $-\mathbf{k}$) and only one type of oscillation pattern appears at different bias voltages. (b) In Bi$_2$Te$_3$, as energy increases, CEC evolves from hexagon, concave hexagon to concave hexagram, and the pairs of EPs are first at ($\mathbf{k}_1$, $-\mathbf{k}_1$); then at ($\mathbf{k}_1$, $-\mathbf{k}_1$), ($\mathbf{k}_2$, $-\mathbf{k}_3$), ($\mathbf{k}_3$, $-\mathbf{k}_2$); then at ($\mathbf{k}_2$, $-\mathbf{k}_3$), ($\mathbf{k}_3$, $-\mathbf{k}_2$); finally at ($\mathbf{k}_4$, $-\mathbf{k}_5$) and ($\mathbf{k}_5$, $-\mathbf{k}_4$). Different types of oscillation pattern appear at different bias voltages.}
\label{fig1}
\end{figure}

Now we turn to the topological SSs on Bi$_2$Te$_3$ and Bi$_2$Se$_3$
with a single Dirac cone near $\bar{\Gamma}$ point in the SBZ on
each surface.  The effective model describing such topological SSs
reads~\cite{fu2009,liu2010}
\begin{equation}
\mathcal{H}(\mathbf{k}) = v(\sigma_xk_y-\sigma_yk_x)+\frac{\lambda}{2}\left(k_+^3+k_-^3\right)\sigma_z,
\end{equation}
where $\hbar\equiv1$, $k_{\pm}\equiv k_y\pm ik_x$, $v$ is the Dirac velocity, $\lambda$ is the warping parameter, and
$\sigma_i$ are Pauli matrices acting on spin space. For simplicity,
here we ignore the particle-hole asymmetry as it affects the shape
of Fermi surface little. The surface band dispersion is
\begin{equation}
\varepsilon_{\pm}\left(k_x,k_y\right) = \pm\sqrt{v^2k^2+\lambda^2k^6\sin^2\left(3\theta\right)},
\end{equation}
where $\varepsilon_{\pm}$ denotes respectively the upper and the
lower energy bands touching at the Dirac point, and $\mathbf{k}
\equiv (k,\theta)$ with $\theta$ as the angle between the wave
vector $\mathbf{k}$ and $k_x$-axis~($\bar{\Gamma}$-$\bar{M}$). The
step edge is always along the close packed $\bar{\Gamma}$-$\bar{K}$
direction. Defining the characteristic energy $\varepsilon^*\equiv
v\sqrt{v/\lambda}$
and length $\sqrt{\lambda/v}$, in Fig.~\ref{fig1} we plot a set of CEC of the upper band in momentum space
for Bi$_2$Se$_3$ and Bi$_2$Te$_3$, respectively. In Bi$_2$Se$_3$
$\lambda=128$eV$\cdot${\AA}$^3$ and $\varepsilon^*=0.59$eV, so that
the CEC is nearly a circle from $0$ to
$0.42\varepsilon^*$~($0.25$eV)~\cite{kuroda2010}. We plot four
representative CEC shown in Fig.~\ref{fig1}(a). When the Fermi
energy increases, the shape of CEC evolves from a
circle~($E_1=0.31\varepsilon^*$), more hexagon-like
($E_2=0.55\varepsilon^*$), hexagon ($E_3=0.7\varepsilon^*$) and to
concave hexagon ($E_4=0.83\varepsilon^*$). In a wide range of energy
only a single pair of EPs exists at $(\mathbf{k},-\mathbf{k})$, so
the characteristic wave vector is always equal to $2\mathbf{k}$ and
$c=2$. In Bi$_2$Te$_3$ the warping effect is stronger with
$\lambda=250$eV$\cdot${\AA}$^3$ and
$\varepsilon^*=0.23$eV~\cite{chen2009}. As shown in
Fig.~\ref{fig1}(b), EPs evolve with the energy as follows: Single
pair of EPs ($\mathbf{k}_1$,$-\mathbf{k}_1$) at
$E_1=0.7\varepsilon^*$; Multiple pairs of EPs
($\mathbf{k}_1$,$-\mathbf{k}_1$), ($\mathbf{k}_2$,$-\mathbf{k}_3$)
and ($\mathbf{k}_3$,$-\mathbf{k}_2$) at
$E_2=1.46\varepsilon^*>E_c\equiv3^{1/3}\sqrt{11/9}\varepsilon^*\simeq1.45\varepsilon^*$;
Two pairs of EPs ($\mathbf{k}_2$,$-\mathbf{k}_3$) and
($\mathbf{k}_3$,$-\mathbf{k}_2$) survive at $E_3=1.91\varepsilon^*$,
as the SSs along the $\bar{\Gamma}$-$\bar{M}$ direction merge into
the bulk conduction band;  No EPs at all at $E_4=2.4\varepsilon^*$,
because the SSs separate from bulk one only in the very vicinity along
$\bar{\Gamma}$-$\bar{K}$ direction on the Fermi surface as observed
in the ARPES experiment~\cite{chen2009}. In this case,
scattering between states around ($\mathbf{k}_4$,$-\mathbf{k}_5$)
and ($\mathbf{k}_5$,$-\mathbf{k}_4$) will be dominant for LDOS
oscillations. Thus in Bi$_2$Te$_3$ the characteristic wave vector and the LDOS oscillation period critically depend on the bias. In most
cases we have parameter $c=2$ except for Fermi energy as high as
$E_4$~($c=1$).

\begin{figure*}[t]
\begin{center}
\includegraphics[width=\textwidth]{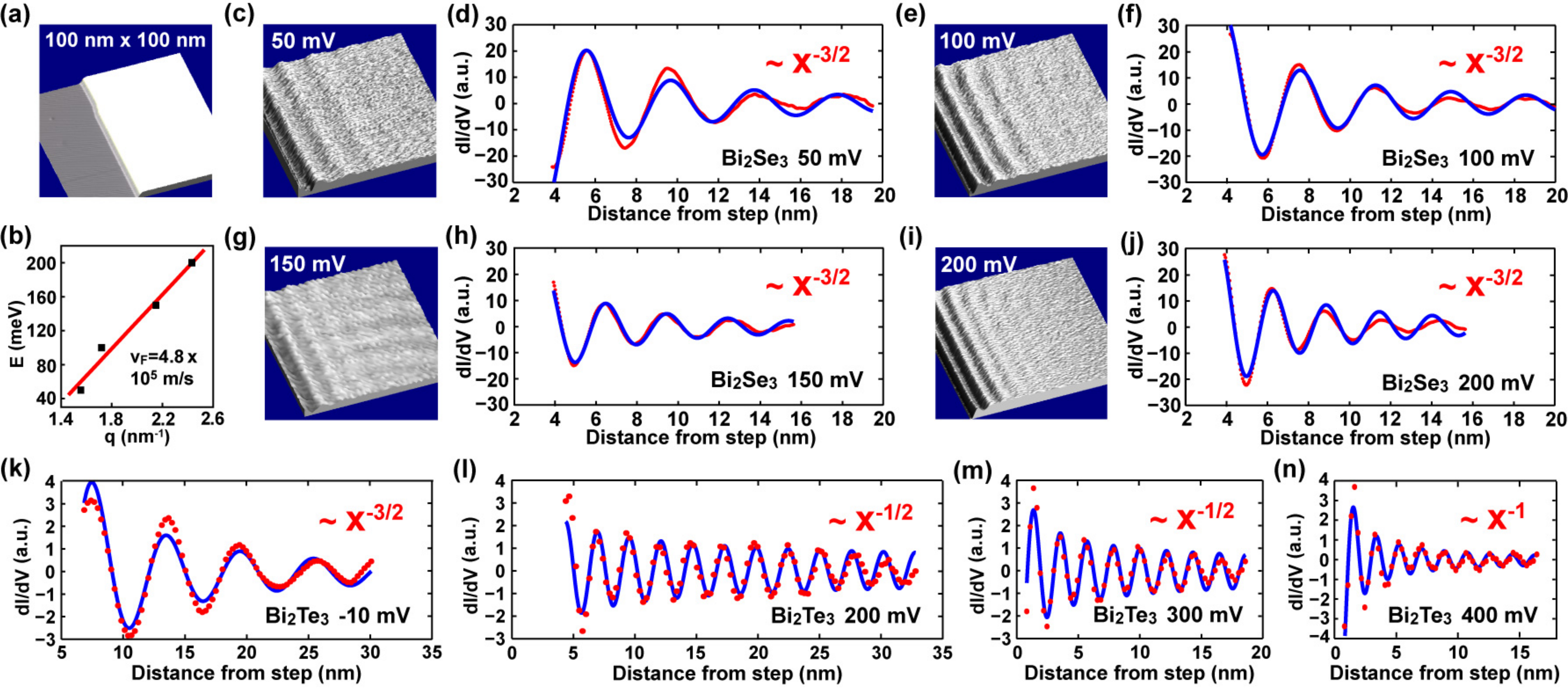}
\end{center}
\caption{(color online) LDOS oscillations due to the step edge
scattering on the surface of Bi$_2$Se$_3$ and Bi$_2$Te$_3$. (a) The
STM topograph of the Bi$_2$Se$_3$(111) film, in which a step edge on
the surface is observed. (b) Energy dispersion deduced from the
standing waves at the step edge. (c)-(j) $dI/dV$ maps and LDOS
oscillations for several bias voltages obtained on the upper terrace
in film shown in (a). (k)-(n) The LDOS oscillations on Bi$_2$Te$_3$
at 4 bias voltages. The LDOS is averaged over the width of the step
and represented by red dots, and the fitting lines are in blue.}
\label{fig2}
\end{figure*}

For the incoming state with wave vector
$\mathbf{k}^i$=$(k^i,\theta^i)$ and energy $\varepsilon_+(k^i)$, the
inner product of two spin wave functions
$\xi_i^{\dag}\xi_f=\sin\theta^i+i\lambda(k^i)^3\sin(3\theta^i)\cos\theta^i/\varepsilon_+(k^i)$.
It is zero only when $\theta^i=0$ as the spins of the time-reversal
pairs $(\mathbf{k},-\mathbf{k})$ are orthogonal. Thus, $b=1$ in
Bi$_2$Se$_3$; while in Bi$_2$Te$_3$, $b=1$ for the pair
($\mathbf{k}_1$,$-\mathbf{k}_1$), and $b=0$ for other pairs of EPs.
Assuming the step edge potential is $V(x)=0$ for $x<0$ and
$V(x)=-V_0$ ($V_0>0$) for $x>0$,  by matching the boundary condition
at the edge the reflection coefficient can be obtained as
\begin{equation}
r(\theta^i) = \frac{e^{-i(\theta^i-\theta^f)/2}-\beta(\theta^i)e^{i(\theta^i-\theta^f)/2}}
{e^{-i(\theta^i+\theta^f)/2}+\beta(\theta^i)e^{i(\theta^i+\theta^f)/2}},
\end{equation}
where $\beta(\theta^i)=(\varepsilon_+(k^i)/k^i+\lambda(k^i)^2\sin(3\theta^i))
/(\varepsilon_+(k^f)/k^f+\lambda(k^f)^2\sin(3\theta^f))$,  $\varepsilon_+(k^i)=\varepsilon_+(k^f)-V_0$ and
$\theta^f(\theta^i)=-\theta^f(-\theta^i)$. Due to the constraint by TRS, $r(\theta^i)=-r(-\theta^i)$, and $r(\theta^i=0)=0$. Thus $a=1$ for $(\mathbf{k},-\mathbf{k})$ pair in Bi$_2$Se$_3$ and
($\mathbf{k}_1$,$-\mathbf{k}_1$) pair in Bi$_2$Te$_3$, and $a=0$ for
other pairs in Bi$_2$Te$_3$.

In short, the algebraical decay index is $3/2$ for
$(\mathbf{k},-\mathbf{k})$ and ($\mathbf{k}_1$,$-\mathbf{k}_1$)
pairs, $1/2$ for ($\mathbf{k}_2$,$-\mathbf{k}_3$) and
($\mathbf{k}_3$,$-\mathbf{k}_2$) pairs, and $1$ for
($\mathbf{k}_4$,$-\mathbf{k}_5$) and
($\mathbf{k}_5$,$-\mathbf{k}_6$) pairs. Therefore, the LDOS
oscillations of the SSs in Bi$_2$Se$_3$ should decay as $x^{-3/2}$
in a wide range of energy (as long as $E<0.85$~eV), much faster than
$x^{-1/2}$ as in 2DES~\cite{crommie1993}. On Bi$_2$Te$_3$
surfaces, as the bias increases, LDOS oscillations decay first as
$x^{-3/2}$ ($E<0.33$~eV), then as a combination of $x^{-3/2}$ and
$x^{-1/2}$, then as $x^{-1/2}$, and finally $x^{-1}$.

To experimentally confirm the above predictions, we analyzed the
interference fringes at the step edges on Bi$_2$Se$_3$ and
Bi$_2$Te$_3$ surfaces. All experiments were carried out at 4.2 K in
an ultrahigh-vacuum low temperature STM system (Unisoku) equipped
with molecular beam epitaxy (MBE) for film growth. The
stoichiometric films of Bi$_2$Se$_3$ and Bi$_2$Te$_3$ were
respectively prepared on graphene and Si(111) substrates by
MBE~\cite{li2010,song2010}. A typical STM image of Bi$_2$Se$_3$
film with a thickness of 50 quintuple layers is shown in
Fig.~\ref{fig2}(a). We can clearly see the atomically flat
morphology of the film and the step of the height of a quintuple layer.
The steps are preferentially oriented along the three close packing
~($\bar{\Gamma}$-$\bar{K}$) directions. The LDOS of electrons at
energy eV is measured through the differential tunneling conductance
$dI/dV$ maps by STS. The Fermi velocity along
$\bar{\Gamma}$-$\bar{M}$ is deduced to be $4.8\times10^5$~m/s by
fitting the linear dispersion curve [Fig.~\ref{fig2}(b)], in good
agreement with the first-principles calculation and the ARPES
data~\cite{zhanghj2009,xia2009}. Figs.~\ref{fig2}(c)-
\ref{fig2}(j) exhibit the $dI/dV$ maps on the upper terrace by the
step shown in Fig.~\ref{fig2}(a) at various bias voltages, and the
LDOS as a function of the distance $x$ from the step. The Dirac
point is at about $0.18$~eV in STS, so the energy of the surface
electron counted from the Dirac point in Fig.~\ref{fig2}(j) is
0.38~eV~(0.64$\varepsilon^*$). The best fit to the LDOS oscillations
is given by $\delta\rho\propto\cos(\Delta k_{x0}x+\phi)x^{-3/2}$ as predicted.
The suppression of backscattering and the circle-like shape of CEC
lead to a much faster decay of LDOS in Bi$_2$Se$_3$ than that in
2DES~\cite{crommie1993}.

\begin{figure}[t]
\begin{center}
\includegraphics[width=2.4in]{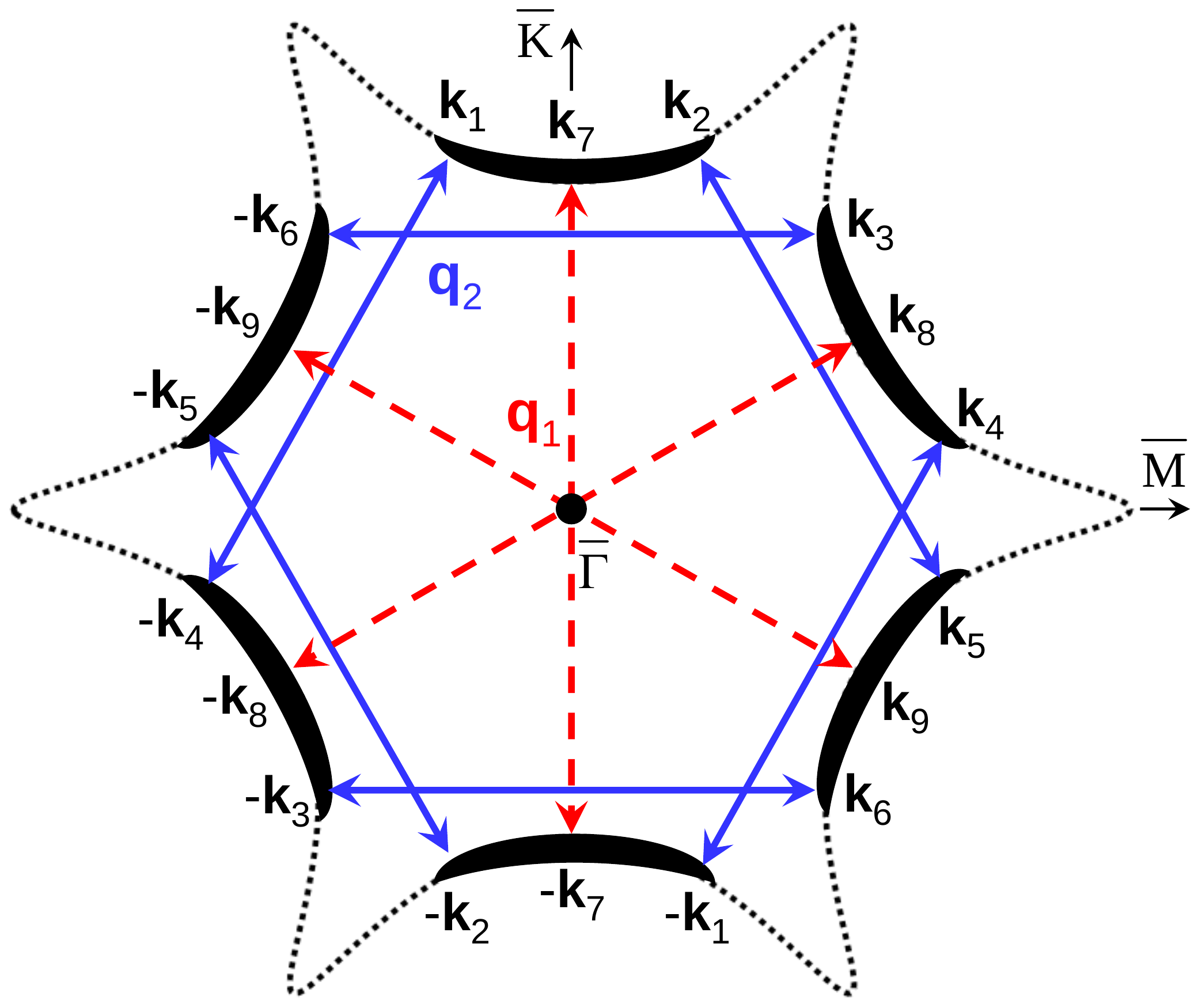}
\end{center}
\caption{(color online) The scattering geometry due to a
non-magnetic impurity on Bi$_2$Te$_3$. When the CEC is a concave
hexagram, the high DOS regions are denoted by bold lines along the
direction of $\bar{\Gamma}$-$\bar{K}$. Two kinds of characteristic
scattering wave vectors include the forbidden $\vec{q}_1$ (dashed
arrow) and the allowed $\vec{q}_2$ (solid arrow) which connects a
pair of EPs.} \label{fig3}
\end{figure}

The case of Bi$_2$Te$_3$ is even more interesting because of the
strong warping effect.  The LDOS oscillations at 4 different bias
voltages on Bi$_2$Te$_3$ film are shown in Figs. \ref{fig2}(k)-(n),
and the corresponding $dI/dV$ maps were reported in
Ref.~\cite{zhangtong2009} already. The Dirac point is estimated
to be at $-0.21$~eV in STS fitting, thus the energies of the surface
electron in Figs.~\ref{fig2}(k)-\ref{fig2}(n) counted from the Dirac
point are respectively 0.20~eV~(0.87$\varepsilon^*$),
0.41~eV~(1.78$\varepsilon^*$), 0.51~eV~(2.18$\varepsilon^*$),
0.61~eV~(2.60$\varepsilon^*$). To extract the decay behavior from
the LDOS data, for each energy we have compared the fittings with
different decay indices. The best fitted LDOS oscillations clearly
demonstrate the crossover of the decay from $x^{-3/2}$ to a
combination of $x^{-3/2}$ and $x^{-1/2}$, then to $x^{-1/2}$, and
then to $x^{-1}$ as the bias voltage increases, which agree well
with our theoretical predictions. Besides,  as predicted theoretically and shown in Figs.~\ref{fig2}(l)-\ref{fig2}(n), the LDOS oscillates with shorter period as increasing the bias.
It is noticeable that due to the strong warping effect of the CEC,
the decay rate of LDOS oscillations on Bi$_2$Te$_3$ is not always
faster than in 2DES.

In the last part, we extend our general formalism for the step edge
scattering to the non-magnetic impurity scattering on Bi$_2$Te$_3$.
Suppose an incident wave $\psi_i\propto e^{i\mathbf{k}^i\cdot\mathbf{r}}/\sqrt{r}$ is scattered into
$\psi_f\propto f(\theta)e^{i\mathbf{k}^f\cdot\mathbf{r}}/\sqrt{r}$
by the impurity potential of $U=U_0\delta(\mathbf{r})$,  then the
LDOS oscillates as $\delta\rho\propto
f(\theta)e^{i(\mathbf{k}^f-\mathbf{k}^i)\cdot\mathbf{r}}/r$ with
$f(\theta)$ denoting the scattering amplitude. The characteristic
wave vector of the QPI pattern at large distance is obtained when
$e^{i(\mathbf{k}^f-\mathbf{k}^i)\cdot\mathbf{r}}$ is stationary for
certain direction $\hat{r}$.
With the concave hexagram Fermi surface of Bi$_2$Te$_3$ as shown in
Fig.~\ref{fig3}, there exist only two kinds of characteristic wave
vectors: $\vec{q}_1$ along $\bar{\Gamma}$-$\bar{K}$ and $\vec{q}_2$
along $\bar{\Gamma}$-$\bar{M}$ direction. Obviously, the $\vec{q}_1$
connects a pair of TRS states whose scattering is forbidden. The
$\vec{q}_2$ connects a pair of states at EP, which dominates the
spatial decay. In such scattering geometry, numerically we find
$q_2$ vary linearly with the energy and $q_2=1.5\bar{k}$ where
$\bar{k}$ is the length of $\bar{\Gamma}$-$\mathbf{k}_7$. Together
with the STM data in Rec.~\cite{zhangtong2009} we obtain the
Dirac velocity along $\bar{\Gamma}$-$\bar{K}$ as
$v=4.15\times10^5$~m/s 
, in good agreement with ARPES result
($v=4.05\times10^5$~m/s)~\cite{chen2009}.
In Bi$_2$Se$_3$, the CEC is circle-like up to $0.25$~eV and
the characteristic scattering wave vector is always along the diameter.
Therefore, we expect the Fourier transformation of LDOS on
Bi$_2$Se$_3$ surface is ring-like.

In conclusion, our theoretical and experimental investigations
indicate that the LDOS oscillation on the surface of TIs is
generally determined by the scattering between surface states around the
extremal points on Fermi surface, either by step edges or by
non-magnetic impurities. We have directly observed different
standing wave patterns caused by scattering off a step on
Bi$_2$Te$_3$ and Bi$_2$Se$_3$ with various warped surface bands,
which, together with the decay indices at different bias voltages,
clearly demonstrate the 2D Dirac nature of topological surface
states.

We thank S.~C.~Zhang, Y.~Y.~Wang, and R.~B.~Liu for helpful discussion. This work is supported by the NSFC Grant No.~11074143, and the Program of Basic Research Development of China Grant No. 2011CB921901.

\end{document}